\documentclass{jfm}
\usepackage{graphicx}
\usepackage{natbib}
\usepackage{epstopdf}
\usepackage{amsmath}
\usepackage{amssymb}
\usepackage{psfrag}
\usepackage{multirow}

\usepackage{color}

\usepackage{enumerate}

\providecommand\bnabla{\boldsymbol{\nabla}}
\providecommand\bcdot{\boldsymbol{\cdot}}

\newsavebox{\astrutbox}
\sbox{\astrutbox}{\rule[-5pt]{0pt}{20pt}}

\newcommand\pd{\partial}
\renewcommand{\vec}[1]{\boldsymbol{\mathrm{#1}}} 
\newcommand{\ten}[1]{\overline{\overline{\boldsymbol{\mathrm{{#1}}}}}}
\newcommand{\pdt}{\partial_t}

\newcommand{\ord}{\epsilon}
\newcommand{\ordest}[1]{\mathcal{O} \left( #1\right)}
\newcommand{\vare}[3][]{ #3^{#1(#2)} }

\newcommand{\Ac}{{\mathcal{A}}}
\newcommand{\ys}{y_s} 

\usepackage{xstring}
\newcommand{\orde}[2]{%
    \IfEqCase{#1}{%
        {0}{  #2^{(#1)} }%
        {1}{ \epsilon #2^{(#1)} }%
    }[ \epsilon^{#1} #2^{(#1)} ]%
}
\newcommand{\ordep}[2]{%
    \IfEqCase{#1}{%
        {0}{  #2^{(#1)} }%
        {1}{ + \epsilon #2^{(#1)} }%
    }[ + \epsilon^{#1} #2^{(#1)} ]%
}
\usepackage{pgffor}

\graphicspath{{img/}}

\title[Boundary conditions at interface between free fluid and a porous medium]
{A framework for computing effective boundary conditions at the interface between free fluid and a porous medium}
\author[U. L\={a}cis, S. Bagheri]%
{U\v{g}is L\={a}cis  and  Shervin Bagheri\thanks{Email address for correspondence: shervin@mech.kth.se}}

\affiliation{Linn\'{e} Flow Centre, Department of Mechanics KTH, SE-100 44 Stockholm, Sweden}

\begin{document}

\maketitle

\begin{abstract}
Interfacial boundary conditions  determined from empirical or \textit{ad-hoc} models remain the standard approach to model fluid flows over porous media, even in situations where the topology of the porous medium is known. We propose a  non-empirical and accurate method to compute the effective  boundary conditions at the interface between a porous surface and an overlying flow. 
Using multiscale expansion (homogenization) approach, we derive a tensorial generalized version of the empirical condition suggested by \citet{beavers1967boundary}. 
The components of the tensors determining the effective slip velocity at the interface are obtained by solving a set of Stokes equations in a small computational domain
near the interface containing both free flow and porous medium.
Using the lid-driven cavity flow with a porous bed, we demonstrate that the derived boundary condition  is accurate and robust by comparing an effective model  to direct numerical simulations.
Finally, we provide an open source code that solves the microscale problems and computes the velocity boundary condition without free parameters over any porous bed.
\end{abstract}

\hrulefill

\section{Introduction}
Surfaces found in nature are generally non-smooth with complex hierarchical  structural features 
 \citep{liu2011bio}. The purpose of these surfaces vary greatly, ranging from camouflage and insulation  to less obvious functions, such as passively interacting with surrounding fluid to reduce drag or noise \citep{abdulbari2013going}. These functions manifest as effective macroscale properties -- for example, permeability, elasticity, slip and optical transparency -- while the origin is the small-scale features of the surface. Therefore, to understand the hydrodynamic function of such complex surfaces, a systematic multi-scale approach is required. In a bottom-up strategy, the microscale fluid-structure physics of the coating material is analysed first; the effective porosity, elasticity or slip are then induced naturally by upscaling the microscale features.

Volume-averaging and homogenization techniques \citep{davit2013homogenization} enable a bottom-up strategy by deriving the effective equations governing the macroscale coating dynamics, which contains parameters arising from microscale features. Whereas these techniques are routinely applied for homogeneous materials (e.g. the interior of a material), their application to inhomogeneous regions (e.g. near interfaces) has not reached the same level of maturity. One example, which is also the focus of the present work, is the interface between an overlying flow and a rigid porous surface. Recent work \citep {ochoa1995momentum_Theory, mikelic2000interface, auriault2010beavers,minale2014momentum} have  treated the inhomogeneous interface problem theoretically with upscaling techniques.  \cite{ochoa1995momentum_Theory} used a volume-averaging technique to derive a shear-stress jump condition. Later on, \cite{valdes2013velocity}  used the same technique to analyse both stress and velocity jump across the interface. Interestingly, they identified a fixed  location of the interface that yields best results when imposing a velocity jump. This is in contrast to both the theoretical findings by \cite{marciniak2012effective} and the numerical results presented in this paper, which show that the accuracy of the velocity jump condition is independent of the interface location. Recently, \cite{minale2014momentum} re-derived the boundary conditions of \cite{ochoa1995momentum_Theory}, elucidating how the stress from the free fluid is partitioned between the porous skeleton and the porous flow. 

Volume-averaging techniques induce closure problems that need to be resolved using scale estimates. Homogenization techniques, on the other hand, begin with scale estimates and an expansion in small parameter $\epsilon=l/H$, defining the scale separation between microscale $l$ and macroscale $H$. With a homogenization approach, one obtains equations at different orders of $\epsilon$ and a decoupling of different quantities and thus also in simpler closure problems. \cite{mikelic2000interface}  used homogenization and method of matched asymptotic expansions to show that the \cite{saffman1971boundary} version of the empirical boundary condition by \cite{beavers1967boundary} (called BJ condition hereafter) is mathematically justified and its slip parameter can be computed by solving microscale problems in an interface unit cell. \cite{auriault2010beavers} also used a homogenization technique to derive a BJ-type of boundary condition valid for pressure-driven flows; he obtained however the condition at different order compared to \cite{mikelic2000interface}, as seen in discussion by \cite{jager2010letter} and \cite{auriault2010reply}. More recently, \cite{carraro2015effective} repeated the  procedure of \cite{mikelic2000interface} to determine the boundary condition of penetration (wall-normal) velocity component.

The boundary conditions derived using upscaling techniques have remained at a proof-of-concept level and only demonstrated on canonical one-dimensional flows. The theoretical progress has not yet resulted in a method that can in a straight-forward manner be applied by practitioners and engineers. The reason is to some extent the non-trivial mathematical aspects -- such as closure problems. Another reason is that the focus has been on mathematically justifying empirical boundary conditions, rather than presenting a  step-by-step method for computing interfacial conditions. Therefore, investigations of practical interest of flows over porous media continue using empirical conditions or conditions with free unknown parameters \citep{han2005transmission,le2006interfacial,rosti2015direct,bottaro2016rigfibre}. Although these conditions provide physical models of the flow over porous media, they are based on lumping all unknown effects into few scalar parameters. This approach requires the support of empirical data  \citep{bottaro2016rigfibre} or extensive computations to cover a large interval of parameters  \citep{rosti2015direct}.

In this work, we provide practitioners the framework to  compute accurate interfacial velocity boundary conditions, instead of empirically determining them. We derive the interface boundary condition for slip velocity using  homogenization  and present the relevant Stokes equations to be solved in a microscale interface unit cell. Our main contribution is to provide a set of simple and numerically feasible microscale problems, which once solved, allows for a robust non-empirical effective interface condition.  Our interface condition can be considered as a generalized version of the BJ condition, since it depends on interface permeability tensor and on the interface velocity strain rate tensor.

This paper is organised as follows. In section~\ref{sec:dns-model-cavity}, using the lid-driven cavity with a porous bed as an example, we compare the velocity field computed from a direct numerical simulation to the field obtained by solving the homogenized (averaged) equations with the interface condition that is proposed in this paper. After that, in sections~\ref{sec:gov-decomp-problem}~to~\ref{sec:eff-intf-cond}, we derive the interface boundary condition. More specifically, in section~\ref{sec:gov-decomp-problem} we decompose the  physical domain into a porous part and free-fluid part and define an interface between the two domains. We then introduce the  equations governing the microscale fluid flow in each part as well as their coupling through continuity of velocity and stress at the interface. In section~\ref{sec:interface-problem}, we use multi-scale expansion to derive the relevant Stokes equations to be solved in a microscale interface unit cell in order determine the effective boundary condition. In section~\ref{sec:eff-intf-cond} we derive the interface conditions by employing a homogenization (averaging) technique and relate the obtained results back to the example presented in section~\ref{sec:dns-model-cavity}.
Finally, in section~\ref{sec:conclusions}, we conclude the paper.

\begin{figure}
  \begin{center}
  \includegraphics[width=0.9\linewidth]{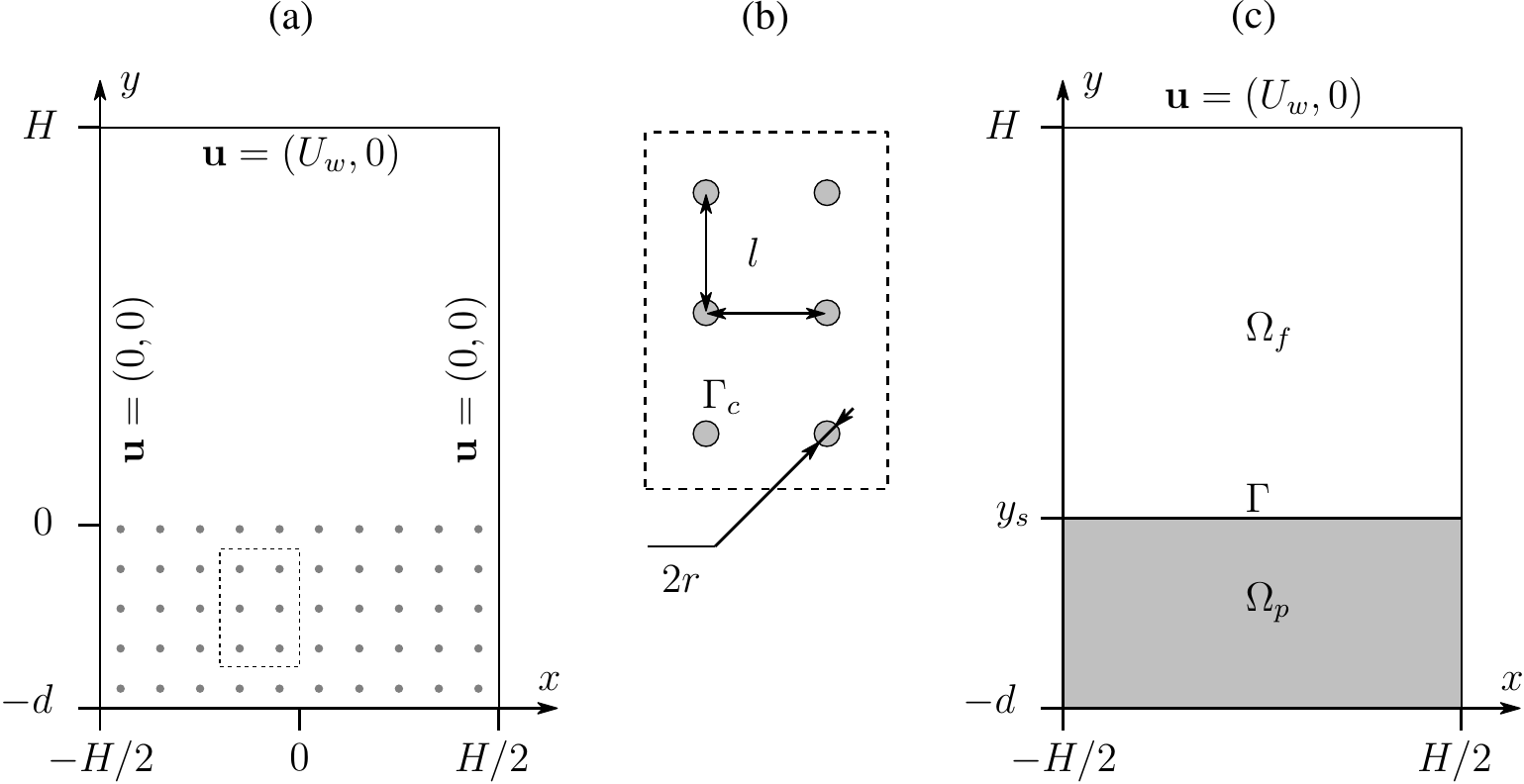} 
  \end{center}
  \caption{
Left frame (a) shows the lid-driven cavity domain with a porous bed. The top wall is driven with velocity $U_w$. Center frame (b) shows a magnified view of the porous bed. Cylinders are spaced apart by distance $l$ and $\Gamma_c$ defines the boundary of the cylinders.  Right frame (c) shows a two-domain description of the same cavity problem in a homogenized setting.
The parameters defining the problem are the volume fraction $\phi = \pi r^2 / l^2 = 0.02$, scale separation $l/H = 0.1$ and porous-bed depth of $d \approx 0.5 H$.}
\label{fig:dns-design-def}
\end{figure}

\section{Direct numerical simulations versus continuum model} \label{sec:dns-model-cavity}
The purpose of this section is to compare two approaches to describe the flow in a lid-driven cavity with a homogenous bed of solid cylinders.  In the first approach, represented in Fig.~\ref{fig:dns-design-def}$a$, we solve the Stokes equations over all spatial scales. This is possible for simplified geometries such as this one, but clearly for more complex (possibly three-dimensional) and much denser porous beds it is not feasible to solve Stokes equations with complete microscale resolution. In the second approach represented in Fig.~\ref{fig:dns-design-def}$c$, we reduce the degrees of freedom of the flow in the porous bed by homogenization.

The cavity has a length $H$ and a depth of $(H+d)$, where the porous bed is confined to $-d<y<0$ and $-H/2<x<H/2$.
Coordinate $y = 0$ corresponds to the tangent plane of the top row of cylinders and the depth of the porous bed is $d \approx H/2$.  The top wall of the cavity is driven by a constant streamwise velocity $U_{w}$, which is sufficiently slow to render fluid inertia negligible inside the cavity. Fig.~\ref{fig:dns-design-def}$b$ shows the geometry of the porous bed, which consists of a lattice of cylinders with diameter $D=2r$ and with the spacing $l$.  For the particular example discussed in this section, the microscale length is $l / H = 0.1$ and the cylinder volume fraction is $\phi = \pi r^2 / l^2  = 0.02$.

\subsection{Direct numerical simulations} \label{sec:dns-desc}
The two-dimensional Stokes equations are solved with no-slip condition imposed at the cylinder surfaces as well as on the vertical and bottom walls of the cavity. The equations are given by,
\begin{align*}
-  \bnabla p + \mu \Delta \vec{u} & = 0, &  &  \\
\bnabla \bcdot \vec{u} & = 0, &  &   \\
\vec{u} & = \left( 0, 0 \right) & \textrm{ on }& \Gamma_n, \ \Gamma_c, \\
\vec{u} & = \left( U_w, 0 \right) & \textrm{ at }& y = H,
\end{align*}
where $\mu$ is fluid viscosity, $\Gamma_n$ denotes the bottom and side boundaries of the cavity, and $\Gamma_c$ denotes the boundary of the cylinders in the porous bed. The computations are performed with FreeFEM++ \citep{MR3043640}, using a triangular mesh and a Taylor-Hood finite element space (P2+P1) for velocity and pressure. We set mesh spacing to $\Delta s_1 = 0.125 l$ at the outer boundaries (cavity walls) of the domain, and $\Delta s_2 = 0.050 l$ at the surface of cylinders\footnote{We have carried out a  simulation of the same configuration with half the mesh spacing ($\Delta s_1 = 0.063 l$ and $\Delta s_2 = 0.025 l$), and observed that the slip velocity changes  by $0.6 \%$.}.

 \begin{figure}
 \begin{center}
 \hspace*{-30pt}
 \includegraphics*[width=.95\linewidth]{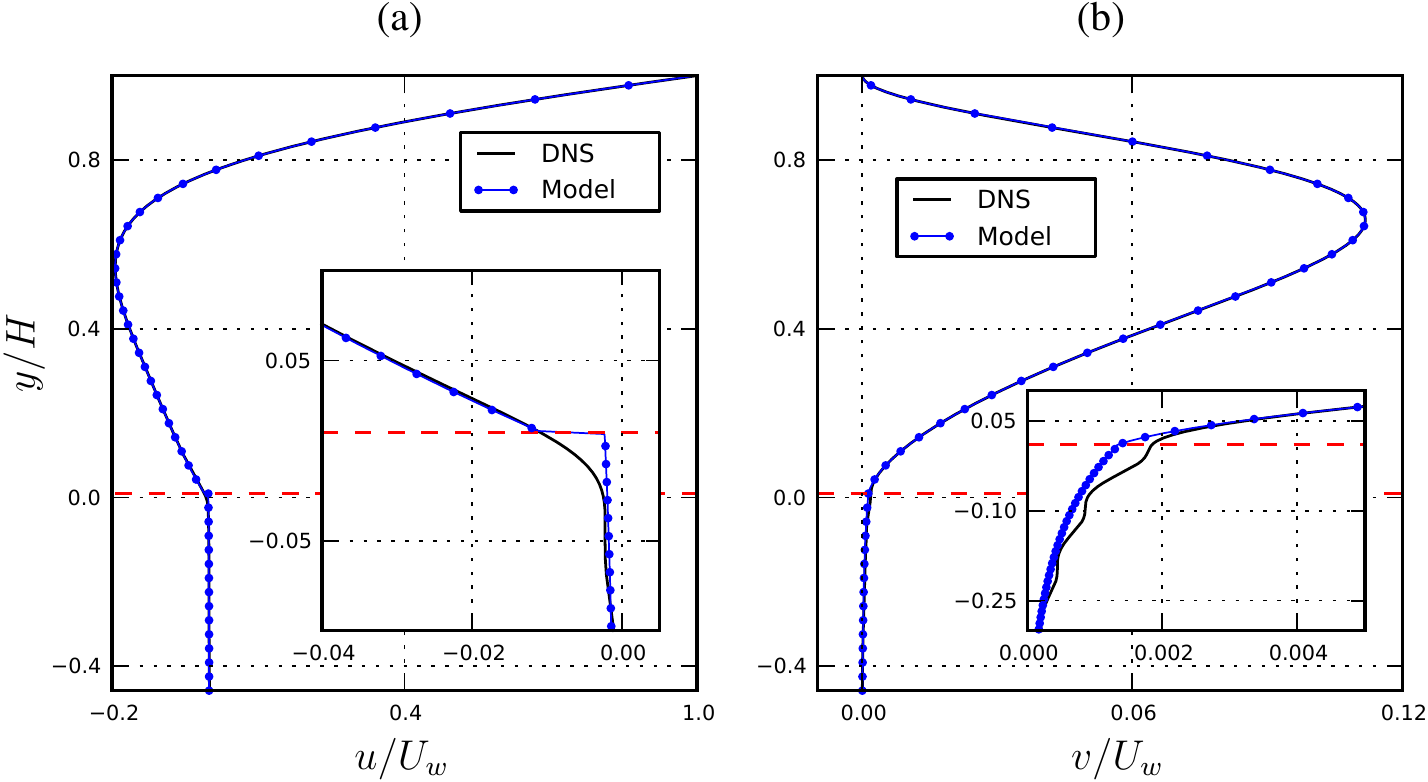} 
 \end{center}
 \caption{
 Left frame (a) shows the streamwise velocity component, whereas the right frame (b) shows wall-normal velocity component at $x/H = -0.1$. Solid black lines depict direct numerical simulations of the porous cavity problem. Blue lines with circular markers correspond to the continuum model of the porous bed coupled to Stokes solver above the bed via interface conditions at $y_s / H = 0.01$ (red dashed line), which is equivalent to $y_s/l=0.1$ in pore scale. Insets show velocity profiles near the interface.}
\label{fig:problem-def}
\end{figure}

\begin{figure}
 \begin{center}
  \hspace*{-30pt}
 \includegraphics*[width=0.6\linewidth]{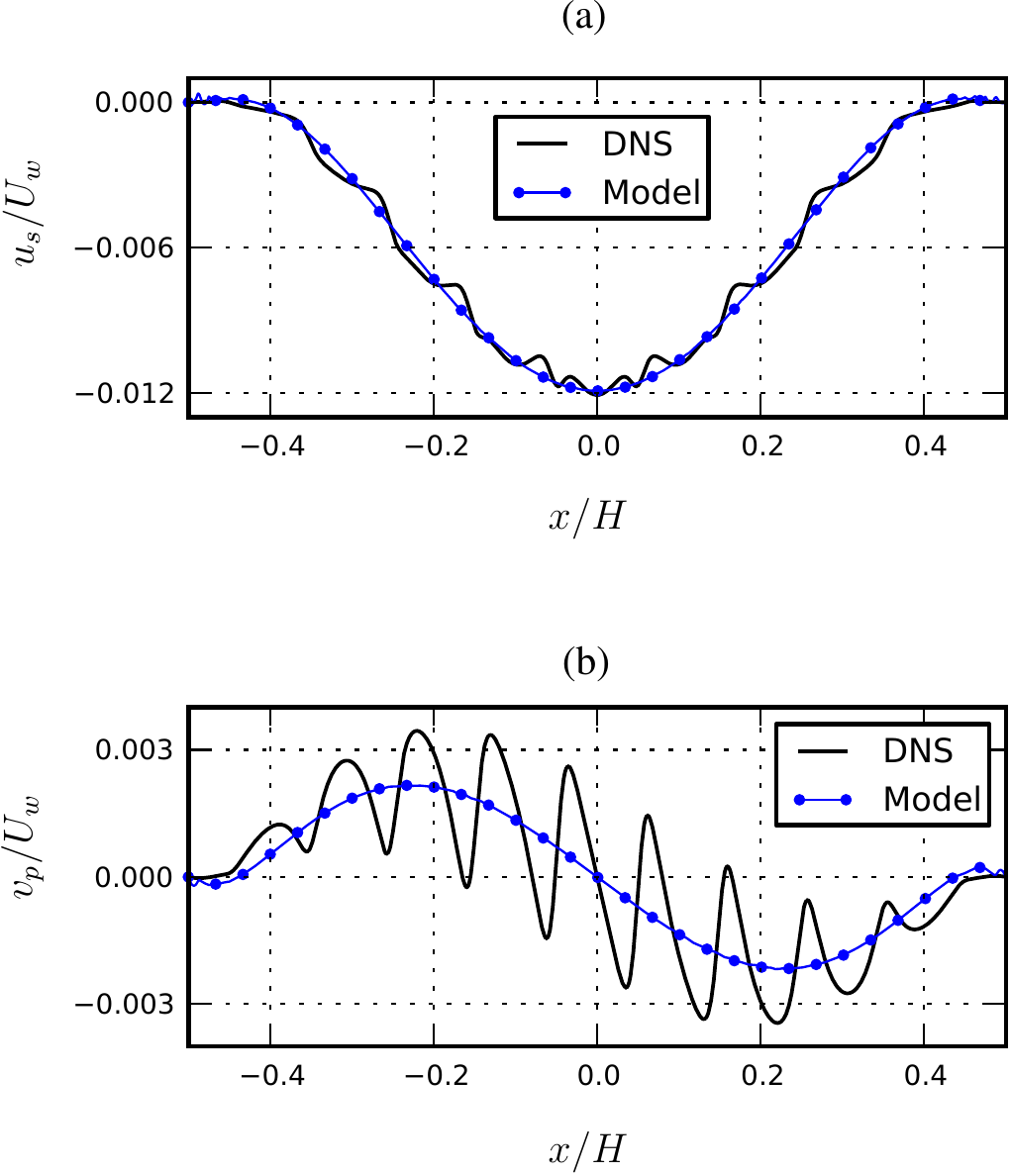} 
 \end{center}
  \caption{In the top frame (a), we compare slip velocity $u_s$ prediction from the continuum model with DNS . The bottom frame (b) compares the penetration velocity $v_p$ between the two approaches. Velocities are sampled at the interface $y = y_s$.}
\label{fig:model-def-results:b}
\end{figure}

In Figs.~\ref{fig:problem-def} and \ref{fig:model-def-results:b}, we present the velocity profiles obtained from DNS with solid black lines. Fig.~\ref{fig:problem-def} shows the streamwise and wall-normal velocity profiles  for the fixed streamwise position $x/H=-0.1$. The insets show the detailed  microscale fluctuations  of the velocities  near the interface and the rapid transition to the macroscale velocity in the free fluid region.  This transition occurs in a thin layer near the top row of cylinders.

Fig.~\ref{fig:model-def-results:b} shows the velocity along $x$ 
at a virtual free-fluid-porous interface placed at $y_s / l =0.1$.  %
If $y_s$ would have been an interface with a rigid wall, these graphs would show the no-slip condition, i.e. zero velocity for both wall-normal and streamwise velocity components. However, at the interface with a porous medium, there is a slip velocity $u_s$ and a penetration velocity $v_p$. The underlying structure of porous medium manifests as microscale oscillations in slip and penetration velocities. Apart from the microscale oscillations, we can observe that both velocity components exhibit macroscale variations. The negative slip velocity, which is induced by the spanwise vortex above the interface, has a maximum value $u_s/U_{w} = -0.0121$ at the center of the cavity.  Although the slip velocity is small compared to the bulk flow, it may have a significant physical effect on the characteristics of the overlying fluid. For example \cite{rosti2015direct} recently showed that slip velocities  below 3\% had a significant effect on the flow statistics in a turbulent channel. Additionally, \cite{carotenuto2013use} showed that precise predictions of slip velocity can be essential to get accurate viscosity measurements from rheology tests. The penetration velocity shows a sinusoidal behaviour; for $x<0$, there is a net mass transport  from the pore region to the free-fluid region, whereas for $x>0$ the net mass flow is in the opposite direction. Similarly, the net momentum transport into the porous region is in opposite directions whether $x>0$ or $x<0$.

The  values of $u_s$ and $v_p$, which are essential to capture the momentum and mass transport across the interface,  depend both on the flow in the pores and on the microscale geometry of the pores. In the next section, we introduce a fully non-empirical method to compute $u_s$ and $v_p$ with an error of $\mathcal{O}{(l/H)}$ without resorting to DNS of the full domain.

\subsection{Simulation of homogenized equations} \label{sec:mod-desc}
We start by replacing the full DNS domain with two rectangular domains, where the free fluid region  $\Omega_f$ and porous region $\Omega_p$ are separated by the interface $\Gamma$, as  shown in Fig.~\ref{fig:dns-design-def}$c$.
In the free fluid part, we do not employ homogenization and therefore the flow is governed by  Stokes equations
\begin{align}
-  \bnabla \hat{p} + \mu \Delta\hat{\vec{u}} & = 0,
\label{eq:sec2:model-gov-1} \\
\bnabla \bcdot \hat{\vec{u}} & = 0,
\label{eq:sec2:model-gov-2}
\end{align}
where $\hat{\vec{u}}$ and $\hat{p}$ are flow and pressure fields in $\Omega_f$, respectively. Dirichlet conditions are imposed for $\hat{\vec{u}}$ on the vertical side walls and the top wall of the cavity (as for the full DNS). The boundary condition at the interface $\Gamma$ in contact with porous region is
\begin{equation}
  \hat{\vec{u}} = \left( \hat{u}_s, \hat{v}_p \right),
 \label{eq:interf}
\end{equation}
where the slip velocity and the penetration velocity depend on the flow in the porous region. In the porous part $\Omega_p$, the flow is governed by  the well-known Darcy's law  
\begin{align*}
\hat{\vec{u}} & = - \frac{\ten{K}^{\textrm{itr}}}{\mu} \cdot \bnabla \hat{p},
\end{align*}
and mass conservation
\begin{align*}
\bnabla \bcdot \hat{\vec{u}} & = 0, 
\end{align*}
where $\ten{K}^{\textrm{itr}}$ is the interior permeability tensor. It is convenient to combine  mass conservation with Darcy's law to arrive with a single equation for the pore-pressure, which -- assuming that permeability tensor $\ten{K}^{\textrm{itr}}$ is constant over space and isotropic -- reads
\begin{align}
\Delta \hat{p} & = 0. \label{eq:sec2:model-gov-3}
\end{align}
We complement this equation with homogenous Neumann conditions on the side walls and the bottom wall, which corresponds to zero transpiration.  At the interface $\Gamma$, we impose Dirichlet condition with the pressure obtained from (\ref{eq:sec2:model-gov-1}-\ref{eq:sec2:model-gov-2}). This continuous pressure condition is valid up to
$\ordest{l/H}$ under the theoretical assumptions done in this paper, which will be discussed in following sections. 
Since the particular porous bed we are investigating is isotropic (see Fig.~\ref{fig:dns-design-def}$a$,$b$), the continuity of  pressure at the  interface is also implied by works of \cite{marciniak2012effective} and \cite{carraro2013pressure}.

Returning to the velocity boundary conditions (\ref{eq:interf}), that are required for solving Stokes system (\ref{eq:sec2:model-gov-1}-\ref{eq:sec2:model-gov-2}), we simply state the conditions of order $\ordest{l/H}$ that will be derived in the next sections (with the final result in equation \ref{eq:model-eq-intf-bc-general-dim}). 
The penetration velocity component $\hat{v}_p$ is  given by
\begin{align}
\hat{v}_p & = - \frac{K_\textrm{cyl}}{\mu} \partial_y \hat{p}^{-} , \label{eq:sec2:effec-v}
\end{align}
where $K_\textrm{cyl}$ is the isotropic permeability of the porous medium consisting of a regular array of circular cylinders. Note that, although pressure is continuous at $\Gamma$ in our case, the pressure gradient is not necessarily continuous; $\partial_y \hat{p}^{-}$ in (\ref{eq:sec2:effec-v}) denotes the pressure gradient when approaching the interface from the porous bed. The condition for $v_p$ can also be obtained from mass conservation for a thin rectangular control volume around $\Gamma$ with periodic streamwise velocity on the vertical sides.
The condition for slip velocity $\hat{u}_s$ is
\begin{align}
\hat{u}_s & = - \frac{K_s}{\mu} \partial_x \hat{p}^{-} + L_s  (\partial_y \hat{u} + \partial_x \hat{v} ) . \label{eq:sec2:effec-u}
\end{align}
This expression
is similar to the condition obtained empirically by Beavers and Joseph, except that $K_s$ is the interface permeability (e.g. $K_s \neq K_{\textrm{cyl}}$), related to a semi-permeable transition layer between the porous medium and the free fluid. Another difference with the BJ condition is that the strain term $\partial_x \hat{v}$ is included in addition to $\partial_y \hat{u}$ \footnote{It has been argued that the term $\partial_x \hat{v}$ should be present for curved boundaries \citep{jones1973low}, but to the authors' knowledge, it has not been derived earlier for flat interfaces, although it has be conjectured to exist by \cite{nield2009beavers}.}. The constant $L_s$ is related to the slip length in the Navier boundary condition.
The constants appearing in boundary conditions (\ref{eq:sec2:effec-v}--\ref{eq:sec2:effec-u}) are provided by microscale simulations in interface cells, described in following sections.
In order to provide an overview of the applicability of the derived boundary
conditions, we summarize here briefly the {\it practical} limits that will be determined both
theoretically and numerically in the remaining part of this paper. These
limits are; (i) moderate scale separation $l/H \leq 0.1$; (ii) restriction on the
Reynolds number based on the seepage velocity ($Re_d \leq 1$); and (iii) restriction on the Reynolds number based on the lid velocity $U_w$ ($Re_f \leq \ord^{-1}$).
The corresponding Reynolds numbers are defined later.


We solve the set of equations (\ref{eq:sec2:model-gov-1}--\ref{eq:sec2:effec-u}) using FreeFEM++ with mesh spacing $\Delta s = 0.125 l$. Fig.~\ref{fig:problem-def} (blue curve with circular symbols) compares the obtained velocity profiles over a vertical slice to the DNS results, where one can observe an excellent agreement between the two.  We note that the effective macroscale behaviour is captured, while underlying oscillations arising from the small-scale characteristics of porous bed are not modelled. The consequence of using  $\ordest{l/H}$ accurate model in the interior (Darcy's law) is that the diffusion process from the free flow to the pore flow (which defines the transition layer of height $\sim l$) is not captured. However, from the perspective of the free fluid, the macroscopic effect of the porous bed is essentially the same using fully resolved DNS and the continuum model.
In Fig.~\ref{fig:model-def-results:b}$a$ (blue curve with circular symbols), we also observe a  good agreement between DNS and the continuum model for slip velocity over a horizontal slice, despite that the latter approach does not resolve the microscale dynamics between and around the cylinders.
The model is able to predict the maximum slip velocity at the center of the cavity $\hat{u}_s/U_{w} = -0.0119$.  Fig.~\ref{fig:model-def-results:b}$b$ compares the predicted penetration velocity and DNS results. There one can observe that, although microscale oscillations dominate the DNS results, the macroscale sinusoidal behaviour is correctly captured by the model.

To close this section, we want to point out that the effective problem is computationally much cheaper than  DNS. The number of degrees of freedom used for the DNS in section~\ref{sec:dns-desc} for region below the interface is around $2.0\cdot 10^5$, whereas for the two-domain approach in section~\ref{sec:mod-desc} it is $1.4\cdot 10^4$. This difference arises from coarser mesh in porous region, as well as reduced number of variables (model equation defines pressure only). In more complex and three-dimensional cases the difference can be significantly larger, and only  averaged models might be computationally feasible to solve numerically.

\section{Governing equations and flow decomposition} \label{sec:gov-decomp-problem}
\begin{figure}
 \begin{center}
 \includegraphics*[width=0.90\linewidth]{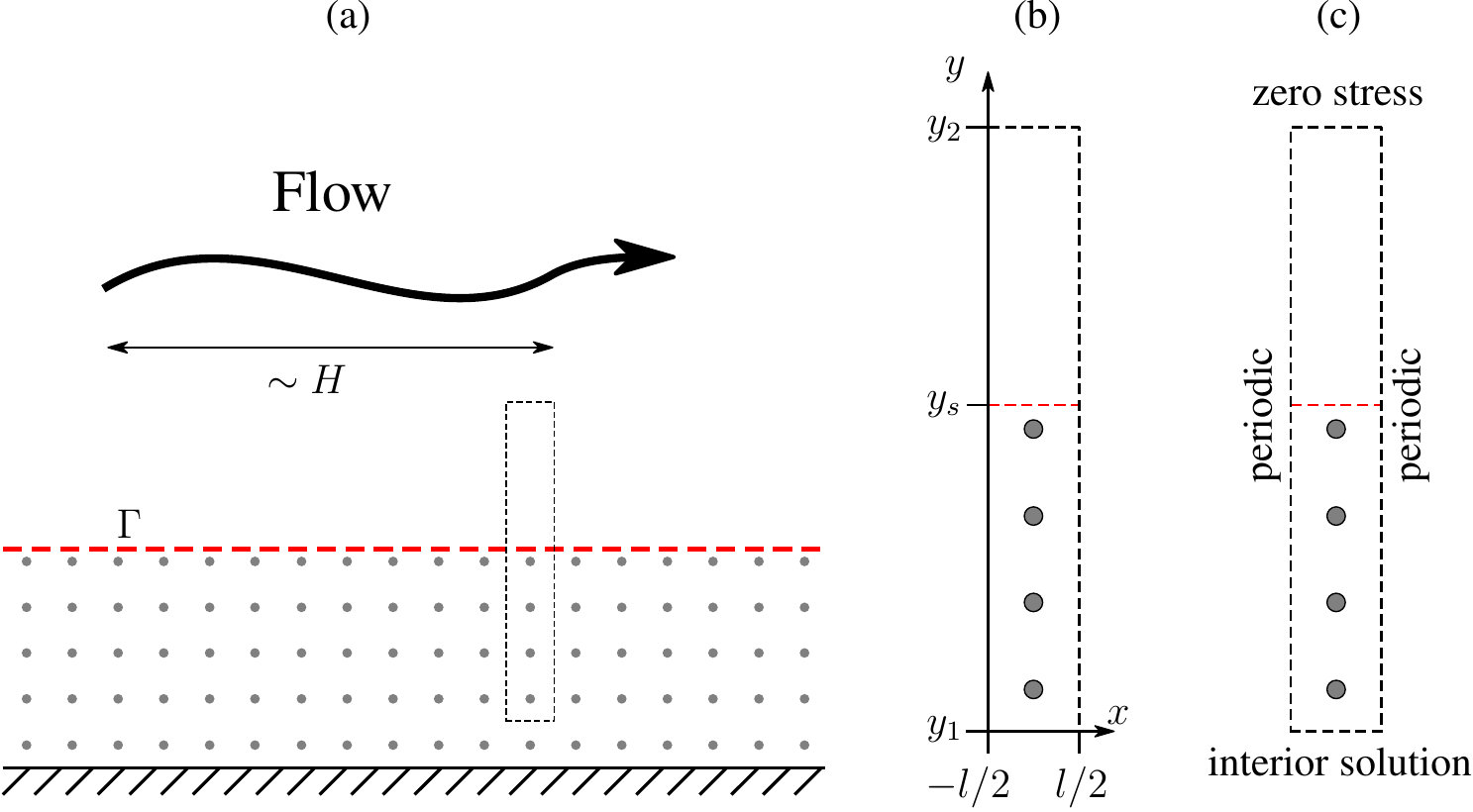} 
 \end{center}
  \caption{Left frame (a) shows a schematic of a flow over porous bed with regular cylinders, where an interface $\Gamma$ between free fluid and porous bed has been introduced. The  dashed rectangular domain corresponds to the interface cell used to compute the effective macroscale boundary condition. The coordinates and the boundary conditions imposed for solving cell problems are shown in  frames (b) and (c), respectively.}
\label{fig:ic-def-cell}
\end{figure}

While the derived effective boundary condition in this paper has been applied to the steady cavity flow over regular array of cylinders, the boundary condition is more general and can be applied to more complex three-dimensional  flows, as schematically shown in Fig.~\ref{fig:ic-def-cell}$a$. We relax the assumptions of a steady and non-inertial flow in the cavity, and start from incompressible Navier-Stokes equations. The length $H$ now corresponds to an appropriate macroscopic length scale of the flow, satisfying $\ord = l/H \ll 1$.

\subsection{Dimensionless Navier-Stokes equations} \label{sec:dimless-NS-equations}

The free fluid region and the porous region are characterized by different spatial and temporal scales. We choose to non-dimensionalize the Navier-Stokes equations using the characteristic scales of the porous medium. In Appendix~\ref{sec:app-scale-estim}, the reader can find detailed analysis of these scales. The  characteristic velocity of the flow in the porous region $U^d$ is
\begin{equation}
U^d \sim \frac{l^2 \Delta P}{\mu H }, \label{eq:first-estim}
\end{equation}
where $\Delta P$ is the characteristic macroscopic global pressure, $\mu$ is the  fluid viscosity, and $H$ and $l$ are the macroscopic and microscopic length scales, respectively. Consequently, we use the following relationships between dimensional (denoted with ``tilde'') and dimensionless variables
\begin{equation}
\tilde{u}_i = U^d u_i, \quad \tilde{p} = \Delta P p, \quad \tilde{x}_i = l x_i, \quad \textrm{and}\quad \tilde{t} = \frac{ l }{ U^d } t.
\label{eq:nondim}
\end{equation}
Here, time is non-dimensionalized with the convection time scale at microscale.
In order to simplify the notation, we use $x_1$ and $x$, and $x_2$ and
$y$ interchangeably. We may now write the Navier-Stokes equations in
the following dimensionless form
\begin{align}
\ord^2 Re_d \left( \pdt u_i  + u_j u_{i,j} \right) & = -p_{,i} + \ord u_{i,jj}, \label{eq:gov-eq-fluid1}\\
u_{i,i} & = 0, \label{eq:gov-eq-fluid2}
\end{align}%
where  $Re_d = \rho_f U^d H / \mu$ is the Darcy Reynolds number. The different order of the scale separation parameter
$\ord$
in front of the terms provides an estimate of the relative magnitude of the terms within the porous region; pressure force plays a dominant role and inertial force is a higher-order effect. This conclusion holds if $Re_d \leq 1$, which is an assumption for the presented approach. Technically, equations (\ref{eq:gov-eq-fluid1}-\ref{eq:gov-eq-fluid2}) hold also in the free fluid region, although the relative magnitude between the terms is not anymore characterized by the scale separation parameter; the scaling (\ref{eq:nondim}), except for pressure, is not suited for the free flow.

\begin{table}
\begin{center}
\begin{tabular}{p{32mm} p{9mm} p{15mm} p{15mm}p{18mm}p{18mm}p{10mm}}
& \hspace{-25mm} Quantity & Scale &  Domain & Microscale dependence & Macroscale dependence & Order \\\hline
Fast flow &$U_i$ & $U^f$ &$\Omega_f$ & No & Yes & $\mathcal{O}{(\epsilon^{-1})}$ \\
Global pressure &$P$ & $\Delta P$ & $\Omega_f$ & No & Yes & $\mathcal{O}{(1)}$ \\
Velocity perturbation &$u_i^+$ & $U^d$ & $\Omega_f$ & Yes & Yes & $\mathcal{O}{(1)}$ \\
Pressure perturbation &$p^+$ & $\Delta p$ & $\Omega_f$ & Yes & Yes & $\mathcal{O}{(\epsilon)}$ \\
Slow flow &$u_i^-$ & $U^d$ & $\Omega_p$ & Yes & Yes & $\mathcal{O}{(1)}$ \\
Pore pressure &$p^-$ & $\Delta P$ &$\Omega_p$ & Yes & Yes & $\mathcal{O}{(1)}$ \\
\end{tabular}
\end{center}
\caption{List of the defined quantities and their properties. For each dimensionless quantity we provide the corresponding dimensional scale, domain, illustrate if microscale and macroscale variations are present, and also state the dimensionless order. Appendix \ref{sec:app-scale-estim} provides more details and discussion on the scaling.}
\label{tab:scales}
\end{table}%

\subsection{Decomposition of the flow field} \label{sec:fast-slow-decomp}
We continue by choosing an interface $\Gamma$ at a vertical coordinate $x_2=\ys$, which divides the fluid domain into a free fluid region and a porous region. The accuracy of the final interface condition does not depend on $y_s$  (up to certain limits)
as proven by \cite{marciniak2012effective}. We confirm this statement numerically in the section~\ref{sec:accuracy}.

We separate the flow above the interface (domain $\Omega_f$) into a fast  flow $(U_i, P)$ and a perturbation $(u^{+}_i, p^{+})$,
\begin{equation}
u_i  = U_i + u^{+}_i,\qquad p  = P + p^{+}. 
\label{eq:goveq-chan-decomp}
\end{equation}
The terms $(u_i^{+},p^{+})$ are generated by the porous medium and will -- as shown below -- be responsible for the induced slip and penetration velocities. The pressure and the velocity below the interface (domain $\Omega_p$), denoted by
\begin{equation}
u_i=u_i^{-}, \qquad p= p^{-}, 
\label{eq:goveq-chan-decomp-slow}
\end{equation}
represent the slow flow and the pressure field in the pores.
Tab.~\ref{tab:scales} summarizes the introduced quantities in $\Omega_f$ and $\Omega_p$. By inserting the decomposition (\ref{eq:goveq-chan-decomp}) and the quantities (\ref{eq:goveq-chan-decomp-slow}) into equations (\ref{eq:gov-eq-fluid1}--\ref{eq:gov-eq-fluid2}) and grouping the different terms, the equations governing the dynamics of the different quantities are obtained. 

\subsubsection{Fast flow}
The \textit{global} pressure $P$ and the \textit{fast} flow $U_i$ are governed by the Navier-Stokes equations with no-slip condition at $\ys$, i.e.,
\begin{align}
\Ac\left(U_{i}, P, \ord \right) =  & \ord^2 Re_d \left( \pdt U_i  + U_j U_{i,j} \right), & & y \geq y_s, \label{eq:gov-eq-fluid1-chfast} \\
U_i = & 0, & & y = y_s. \label{eq:gov-eq-fluid1-chfast-bc}
\end{align}
Here, $\Ac$ is a linear Stokes operator defined by equation (\ref{app-eq:def-Stokes}) in  Appendix \ref{sec:app-def}, where we have summarized definitions of various operators and tensors. The macroscale pressure associated with fast flow is assumed to have a magnitude of $\tilde{P} \sim \Delta P$. The fast velocity $U_i$ has a characteristic free flow velocity, $\tilde{U}_i \sim U^f$,  that is ``faster'' than the Darcy velocity scale $U^d$; we assume
\begin{equation}
U^f \sim \ord^{-1} U^d. \label{eq:assumption-on-Uf}
\end{equation}
Using the introduced non-dimensionalization (\ref{eq:nondim}) on the variable estimates, we arrive with a priori orders of the fast flow velocity and the global pressure
\begin{align}
 U_i = \ordest{\ord^{-1}},   &&     {P} = \ordest{1}. \label{eq:fastflowscal}
\end{align}
The dimensionless fast flow field $U_i$ becomes very large for small $\epsilon$, whereas the global pressure in $\Omega_f$ -- either externally imposed as in pressure-driven channel flow or induced by the fast flow as in lid-driven cavity -- is of order one. 

Note that the assumption (\ref{eq:assumption-on-Uf}) is not a universal one, and depends on the bulk Reynolds number; for example, for Stokes flow $U^f \sim \ord^{-2} U^d$.  Appendix~\ref{sec:app-scale-estim} shows that  (\ref{eq:assumption-on-Uf}) is obtained, when $Re_f =\rho_f HU^f/\mu\sim \epsilon^{-1}$. This a priori scale estimate simplifies the multiscale expansion outlined in section \ref{sec:interface-problem} and its consequence is of a theoretical nature; in practice, our derived boundary condition predicts very accurately the slip and penetration velocity for a wide range of parameters, as we demonstrate in section \ref{sec:accuracy}.

\subsubsection{Perturbations and slow flow}
The perturbations above the interface are governed by
\begin{align}
\Ac\left(u^{+}_{i}, p^{+}, \ord \right) & = \ord^2 Re_d \left( \pdt u^{+}_i  + u^{+}_j u^{+}_{i,j} + U_j u^{+}_{i,j} + u^{+}_j U_{i,j} \right), & & y \geq y_s, \label{eq:gov-eq-fluid1-chslow} \\
u^{+}_{i} & = u^{-}_{i}, & & y = y_s. \label{eq:gov-eq-fluid1-chslow-bc}
\end{align}
In order to solve these equations, one has to know the flow below the interface ($u_i^-$). 
The pressure and the slow velocity below the interface are governed by
\begin{align}
\Ac\left(u^{-}_{i}, p^{-}, \ord \right) & = \ord^2 Re_d \left( \pdt u^{-}_i  + u^{-}_j u^{-}_{i,j} \right), & & y \leq y_s, \label{eq:gov-eq-fluid1-pslow} \\
\Sigma^{u^-}_{ij} n_j  & = \Sigma^{u^+}_{ij} n_j + \Sigma^{U}_{ij} n_j, & & y = y_s, \label{eq:gov-eq-fluid1-pslow-bc}
\end{align}
where the stress tensors containing $u^+_i$, $u^-_i$ and $U_i$ in (\ref{eq:gov-eq-fluid1-pslow-bc}) are defined by equations (\ref{app-eq:def-stress1}--\ref{app-eq:def-stress3}) in  Appendix \ref{sec:app-def}.
Note that the decomposition of the flow introduced in section \ref{sec:fast-slow-decomp} is exact, since continuity of both velocity (\ref{eq:gov-eq-fluid1-chslow-bc}) and total stress (\ref{eq:gov-eq-fluid1-pslow-bc}) is imposed at the interior interface $\Gamma$. In other words, if one would sum-up  equations (\ref{eq:gov-eq-fluid1-chfast},\ref{eq:gov-eq-fluid1-chslow},\ref{eq:gov-eq-fluid1-pslow}) and the boundary conditions (\ref{eq:gov-eq-fluid1-chfast-bc},\ref{eq:gov-eq-fluid1-chslow-bc},\ref{eq:gov-eq-fluid1-pslow-bc}), the  Navier-Stokes equations (\ref{eq:gov-eq-fluid2}) defined in the full domain  would be recovered.

The perturbation terms $(u_i^{+},p^{+})$ in $\Omega_f$ are an effect of the porous medium. Therefore perturbation velocity is estimated by the seepage velocity, in dimensional setting $\tilde{u}_i^+\sim U^d$, and pressure perturbation by the microscale pressure $\tilde{p}^+\sim \Delta p$, where $\Delta p$ is the pressure induced by the seepage velocity, see Appendix~\ref{sec:app-scale-estim}. In the current non-dimensional setting, we have
\begin{align}
     u_i^{+} &= \ordest{1}, &      p^{+} & = \ordest{\ord}. \label{eq:press-pert-scale-above}
\end{align}
Comparing to (\ref{eq:fastflowscal}), we observe $\| p^{+} \|\ll \| P \|$ and $\| u^{+}_i \|\ll \| U_i \|$
when  $\epsilon \ll 1$.

The slow velocity below the interface can be estimated  as $\tilde{u}^{-}_i  \sim {U^d}$, since it is reasonable to assume that the flow inside porous medium has the magnitude of Darcy velocity.
The pore pressure, on the other hand, can be estimated as $\tilde{p}^{-} \sim {\Delta P}$, because the global macroscale pressure is present also in porous medium. For dimensionless variables we then have
\begin{align}
u^{-}_i & =\ordest{1}, & p^{-} & = \ordest{1}. \label{eq:press-pert-scale-below}
\end{align}
The dimensional estimates and dimensionless orders of the decomposed quantities are summarized in Tab.~\ref{tab:scales}.

\section{Multi-scale expansion} \label{sec:interface-problem}
We now turn our attention to the multi-scale analysis of the flow near the interface and construct an approximate description of it within an interface cell (defined below). To carry out multi-scale expansion, we introduce the macroscale and microscale coordinates 
\[
X_i=\frac{\tilde{x}_i}{H} \qquad \textrm{and} \qquad x_i=\frac{\tilde{x}_i}{l},
\] 
respectively. These coordinates are appropriate to describe the macroscopic and microscopic variations and are related to each other by $X_i=\epsilon x_i$.
In the new coordinates, there are two derivatives appearing due to the chain rule
\begin{equation}
\left( \right)_{,i} = \left( \right)_{,i_1} + \ord \left( \right)_{,i_0}, \label{eq:mse-chain-der}
\end{equation}
where $\left( \right)_{,i_0}$ denotes the derivative with respect to  $X_i$ and $\left( \right)_{,i_1}$ with respect to  $x_i$. 

The fast flow $U_i$� and the global pressure $P$ do not depend on microscale coordiate, i.e. $U_{i,j_1} = 0$ and $P_{,i_1} = 0$. This  is a direct consequence of the definition of fast flow problem (\ref{eq:gov-eq-fluid1-chfast}--\ref{eq:gov-eq-fluid1-chfast-bc}) and is valid for $\ord \ll 1$. For $u_i^\pm$ and $p^\pm$, which depend on both coordinates, we  carry out the multi-scale expansion as explained by \cite{mei2010homogenization}. The perturbation velocity and the pressure above the interface ($y\geq y_s$) are expanded as
\begin{align}
&u^{+}(X_i,x_i) = u_i^{+ (0)}(X_i,x_i) +\ord u_i^{+(1)} (X_i,x_i) + \mathcal{O}(\ord^2),\label{eq:ex11}\\
&p^{+}(X_i,x_i) = \ord p^{+(1)} (X_i,x_i) + \ord^2 p^{+(2)} (X_i,x_i) + \mathcal{O}(\ord^3).\label{eq:ex12}
\end{align}
The pressure expansion starts with $\mathcal{O}(\epsilon)$ term, since  $p^{+} = \ordest{\ord}$.
Below the interface ($y\leq y_s$) the slow flow and the pore pressure are expanded as, %
\begin{align}
&u^{-}(X_i,x_i) =  u_i^{- (0)}(X_i,x_i) +\ord u_i^{-(1)} (X_i,x_i) + \mathcal{O}(\ord^2),\label{eq:ex21}\\
&p^{-}(X_i,x_i) =  p^{-(0)} (X_i,x_i) + \ord p^{-(1)} (X_i,x_i) + \mathcal{O}(\ord^2).\label{eq:ex22}
\end{align}
We insert expansions (\ref{eq:ex11})-(\ref{eq:ex22}) into the corresponding equations (\ref{eq:gov-eq-fluid1-chslow})-(\ref{eq:gov-eq-fluid1-pslow-bc}), and collect the terms at different orders. In the following subsections, we introduce and solve equation systems appearing at first two orders ($\mathcal{O}(1)$ and $\mathcal{O}(\epsilon)$). 

\subsection{ $\mathcal{O}(1)$ equation and its analytical solution in an interface cell} \label{sec:mse-lead-ord-eq}
Collecting the terms with pre-factor $1$, we get the following system
\begin{align}
p^{-(0)} n_i & = P n_i, & & y = y_s, \label{eq:gov-mse-ord0-start} \\
p^{-(0)}_{,i_1} & = 0,  & & y \leq y_s. \label{eq:gov-mse-ord0-end}
\end{align}
We observe that  the zeroth-order pressure in the porous region $p^{-(0)}$ is independent of the microscale coordinate $x_i$. 
For our purpose, which is to derive the macroscale effective boundary condition, it is sufficient to solve  this equation in an elongated cell near the vicinity of the interface (Fig.~\ref{fig:ic-def-cell}$b)$. The size of this cell 
\[
\Omega_{\textrm{cell}}= \{y_1\leq y \leq y_2, -\frac{1}{2}\leq x\leq \frac{1}{2}\}
\]  is chosen in such a way to capture only the microscale behavior near the interface. The solution to (\ref{eq:gov-mse-ord0-start}-\ref{eq:gov-mse-ord0-end}) below the interface in $\Omega_{\textrm{cell}}$ is  constant and equal to global pressure $P$ at the interface,
\begin{align}
 p^{-(0)} & = \left. P \right|_{y_s}, & & y_1\leq y \leq y_s.
 \label{eq:pressure-zero-sol-intf}
\end{align}
We will use this result in section \ref{sec:eff_press} to derive a macroscale pressure condition at the interface.

\subsection{ $\mathcal{O}(\epsilon)$ equation and its computational solution in an interface cell} \label{sec:mse-first-ord-eq}
Next, we collect the first-order terms with pre-factor $\epsilon$, which results in  Stokes equations for $(u_i^{\pm(0)}, p^{\pm (1)}$). Specifically, above the interface, we have
\begin{align}
\Ac_1\left( \vare[+]{0}{u_{i}}, \vare[+]{1}{p}, 1\right)  =  0,  & & y \geq y_s, \label{eq:gov-microcell-fluid0}
\end{align} 
where $\Ac_1$  denotes the Stokes operator, which contains derivatives with respect to microscale $x_i$, as defined in equation (\ref{app-eq:def-Stokes-micro}).
Below the interface, we have
\begin{align}
\Ac_1\left( \vare[-]{0}{u_{i}}, \vare[-]{1}{p}, 1\right)  = p^{-(0)}_{,i_0}(X_i), & & y \leq y_s. \label{eq:gov-microcell-fluid1}
\end{align}
We observe that the slow flow $u_i^-{(0)}$ is forced by the macroscale gradient of the pore pressure term ($p_{,i_0}^{-(0)})$.
In contrast, above the interface the equation for perturbation $u^{+(0)}_i$ is not driven by a lower-order pressure term; the $\mathcal{O}(1)$ pressure above the surface is $P$, which is contained in equations for $U_i$ (\ref{eq:gov-eq-fluid1-chfast}--\ref{eq:gov-eq-fluid1-chfast-bc}). The same global macroscale pressure is driving the fast flow and the slow flow, but whereas this pressure is {\textit{defined}} as $P$ above the interface, it is {\textit{obtained}}  as the leading-order expansion term below the interface. This is a consequence of the fact that the fast flow is not expanded, since it varies only with macroscale coordinate by definition.

The boundary conditions of $\ordest{\ord}$ equations at the interface $\ys$ are continuity of velocity
\begin{align}
\vare[-]{0}{u_{i}}  &=  \vare[+]{0}{u_{i}},  \label{eq:gov-microcell-fluid3}
\end{align}
and jump in stress
\begin{align}
 \vare[u^-]{1}{\Sigma_{ij}}  n_j &= \vare[u^+]{1}{\Sigma_{ij}}  n_j + S_{ij} n_j.  \label{eq:gov-microcell-fluid4}
\end{align}
Here, $\vare[u^\pm]{1}{\Sigma_{ij}}=-\vare[\pm]{1}{p}\delta_{ij}+(\vare[\pm]{0}{u_{i,j_1}}+\vare[\pm]{0}{u_{j,i_1}})$ is the stress tensor of the perturbation velocity and the slow velocity, whereas $S_{ij} = U_{i,j}+U_{j,i}$ is the strain tensor of the fast flow.

The system of equations (\ref{eq:gov-microcell-fluid0}--\ref{eq:gov-microcell-fluid4}) is solved in  the interface cell $\Omega_{\textrm{cell}}$ shown in Fig.~\ref{fig:ic-def-cell}$b$.
To complete the problem formulation, boundary conditions are needed at the sides of the cell. Due to regularity of the porous structure, we impose periodic conditions on the sides, as shown in Fig.~\ref{fig:ic-def-cell}$c$. At the bottom of the cell, we impose the interior solution, which is
\begin{align}
\vare[-]{0}{u_{i}} & = - K^{\textrm{itr}}_{ij} p^{-(0)}_{,j_0}, & & y = y_1, \label{eq:microsc-inter-ansatz}
\end{align}
where $K^{\textrm{itr}}_{ij}$ is the classical interior permeability field. For the derivation of this expression and the corresponding microscale problem, reader is referred to book by \cite{mei2010homogenization}.
From the literature \citep{mikelic2000interface,jager2009modeling,
marciniak2012effective,carraro2013pressure} it is well known that the interface
cell is exposed to
a zero-stress condition at the infinity within the method of matched asymptotic
expansion, therefore we impose zero-stress condition at $y = y_2$.

We may consider the  Stokes equations (\ref{eq:gov-microcell-fluid0}--\ref{eq:gov-microcell-fluid1}) and the boundary conditions (\ref{eq:gov-microcell-fluid3}--\ref{eq:microsc-inter-ansatz}) as one linear problem with four unknowns $(u_i^{\pm(0)}, p^{\pm (1)}$). 
Due to linearity, we can construct the solutions for the velocity and pressure fields as superposition of  $p^{-(0)}_{,i_0}(X_i)$ and $S_{ij} (X_i)$, i.e.
\begin{equation}
\vare[\pm]{0}{u_i} = - K^{\pm}_{ij} p^{-(0)}_{,j_0}+ L^{\pm}_{ijk} \left. S_{jk} \right|_{\ys},
 \label{eq:microsc-ansatz}
\end{equation}
and
\begin{equation}
\vare[\pm]{1}{p} = - A^{\pm}_j p^{-(0)}_{,j_0}+ B^{\pm}_{ij} \left. S_{ij} \right|_{\ys}.
\label{eq:microsc-ansatz1}
\end{equation}
The average of $K_{ij}$ is a tensorial effective Darcy permeability and the average of $L_{ijk}$ is related to the tensorial version of the Navier slip coefficient. Similarly, the tensors $A_j$ and $B_{ij}$ are transfer coefficients from the driving pressure gradient and fluid strain, respectively, to the perturbation pressure.

\begin{figure*}
  \begin{center}
  \hspace*{-30pt}
\includegraphics[width=1.1\linewidth]{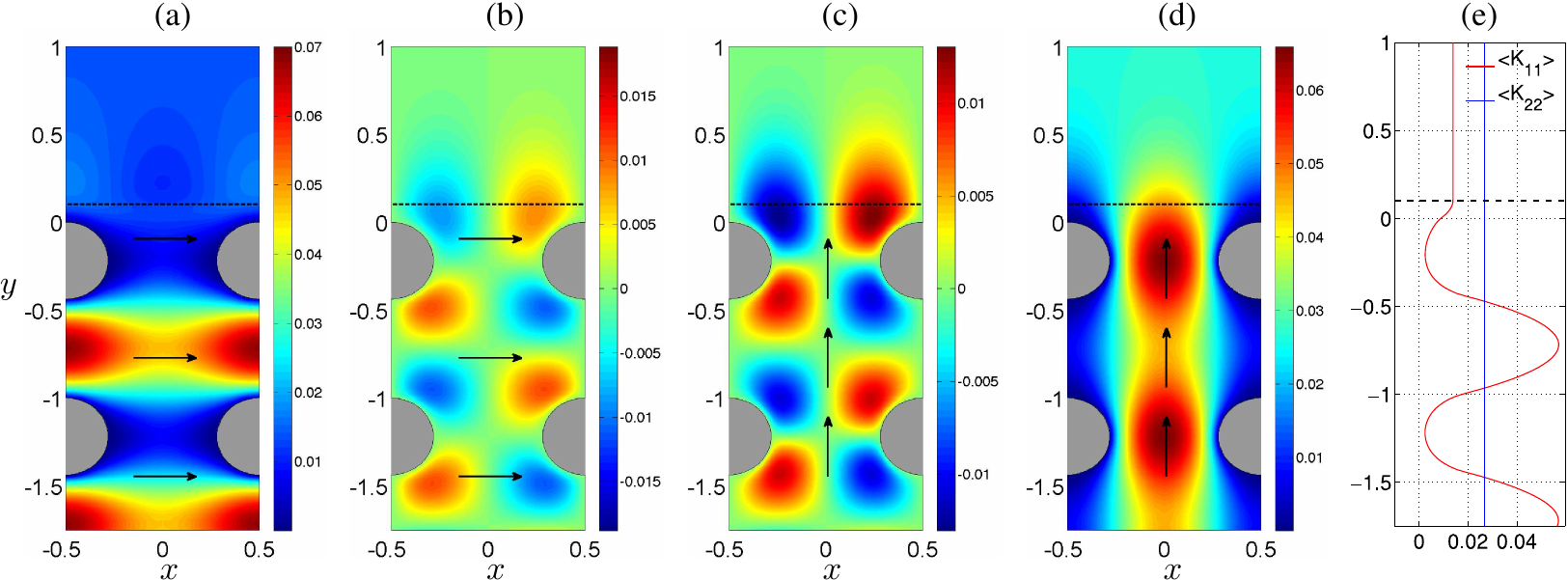}
  \end{center}
  \vspace{0.1cm}
  \caption{Solutions of interface problems for the coefficients of the Darcy term ($\phi = 0.15$ and $y_s=0.1$). The frames from left to right correspond to the flow fields $K_{11},K_{21},K_{12}$ and $K_{22}$. The arrows indicate the direction of the constant unit volume forcing below the interface (horizontal dashed line).  Rightmost frame (e) shows  plane-averaged profiles; the streamwise component provides the interface permeability ($\langle K_{11}\rangle=0.014$ above $y_s$), whereas the wall-normal component is constant and corresponds to  the interior permeability ($\langle K_{22}\rangle=0.0266$).}
\label{fig:ms-res-prof}
\end{figure*}

\subsubsection{Microscale Stokes problems for Darcy term}
By inserting the ansatzes (\ref{eq:microsc-ansatz}-\ref{eq:microsc-ansatz1}) into equations (\ref{eq:gov-microcell-fluid0}--\ref{eq:microsc-inter-ansatz}), it follows that the tensors $K^{\pm}_{ij}$ and $A^{\pm}_i$ satisfy,
\begin{align*}
&\Ac_1\left( K^{+}_{ik}, A^{+}_{k}, 1 \right)   = 0, &\qquad y\geq y_s, \\
&\Ac_1\left( K^{-}_{ik}, A^{-}_{k}, 1 \right)  = -\delta_{ik}, & \qquad y\leq y_s, 
\end{align*}
with boundary conditions at the interface $y_s$ given by
\begin{align*}
 K^{-}_{ik}  & =  K^{+}_{ik},  &  \Sigma^{K^{+}}_{ij}  n_j & =  \Sigma^{K^{-}}_{ij}  n_j.
\end{align*}
At the bottom boundary $y = y_1$, we have $ K^{-}_{ik} = K^{\textrm{itr}}_{ik}$.
The field $K^{\pm}_{jk}$ represents the $j$th velocity component of the $k$th Stokes problem. Thus to determine every component of   $K^{\pm}_{ij}$ and $A^{\pm}_i$, 3  pairs of Stokes problems have to be solved coupled at the interface through continuity of velocity field and stress. Note that below the interface, the flow is driven by a unit forcing in one direction at a time. Therefore the physical interpretation of $K_{i1}$ for example, is the flow response to forcing in the horizontal direction below the interface, as shown in Fig.~\ref{fig:ms-res-prof}$a$,$b$.

To obtain reliable results, the interface cell needs to extend sufficiently into the free fluid such that variations of $K^{+}_{ij}$ are small and sufficiently into the porous medium such that variations of $K^{-}_{ij}$ are periodic. We have investigated different heights of the interface cell, and have determined that height of $10 l$ (containing 5 cylinders below the interface) is sufficient. Fig.~\ref{fig:ms-res-prof} shows $K^{\pm}_{ij}$ fields and corresponding plane averaged profiles near the tip of the solid structure for interface location $y_s = 0.1$.

\subsubsection{Microscale Stokes problems for Navier-slip term}
By inserting the ansatzs (\ref{eq:microsc-ansatz}--\ref{eq:microsc-ansatz1}) into (\ref{eq:gov-microcell-fluid0}--\ref{eq:microsc-inter-ansatz}), the following equations for the tensors $L^{\pm}_{ijk}$ and $B^{\pm}_{ij}$ are obtained,
\begin{align*}
& \Ac_1\left( L^{+}_{ikl}, B^{+}_{kl}, 1 \right)  = 0, & y\geq y_s,\\  
& \Ac_1\left( L^{-}_{ikl}, B^{-}_{kl}, 1 \right) = 0, & y\leq y_s,
 \end{align*}
 with boundary conditions at $y_s$ given by,
\begin{align*}
 L^{-}_{ikl} & = L^{+}_{ikl},   & \Sigma^{L^{+}}_{ij} n_j & = \Sigma^{L^{-}}_{ij} n_j - \delta_{ik} n_l.
 \end{align*}
At the bottom boundary $y = y_1$, we have $ L^{-}_{ikl} = 0$.
The tensors transferring the stress of the free fluid to the perturbation velocity require the solution of $9$ pairs of coupled Stokes problems. The forcing for these equations is at the interface in the form of a stress condition. For example, the $L_{i12}$ component is the flow response to a unit tangental stress at the interface, whereas the $L_{i22}$ is the response to unit normal stress at the interface. In general, for problems with flat interfaces described in a coordinate system aligned with the interface, only three pairs of  problems are forced. 

\begin{figure}
  \begin{center}
        \includegraphics[width=0.7\linewidth]{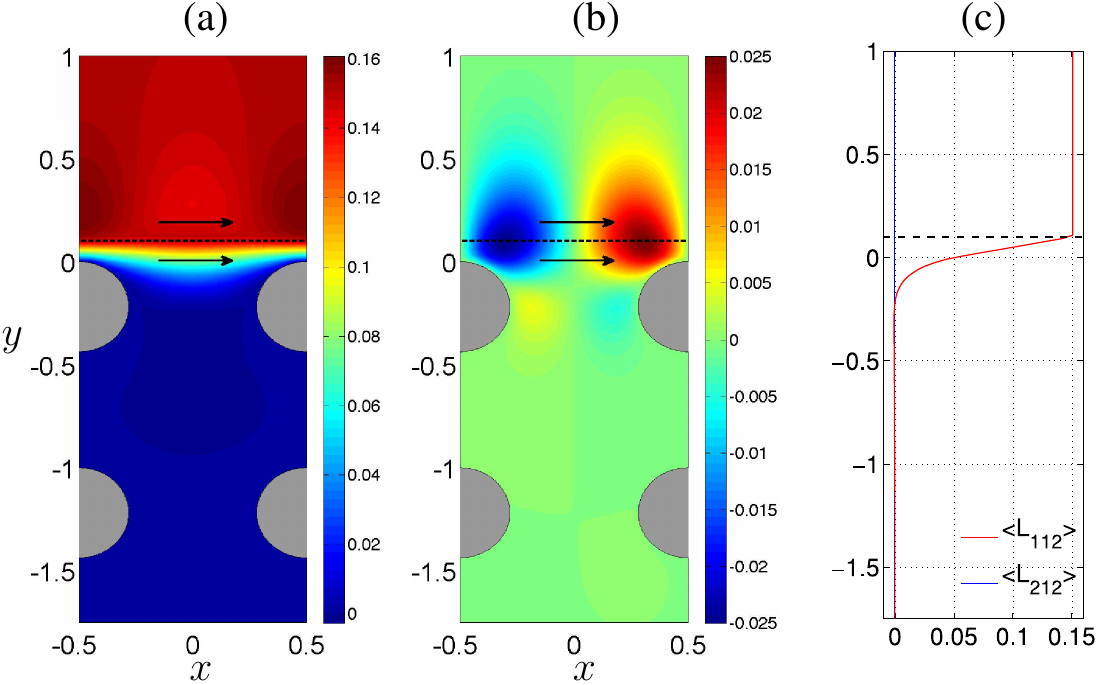}
  \end{center}
  \caption{
  Solutions of interface problem for the non-zero component ($L_{i12}$) of the Navier-slip term ($\phi = 0.15$ and $y_s=0.1$). The left and center frames  correspond to  $L_{112}$ and $L_{212}$, respectively. The arrows indicate the direction of the constant unit boundary forcing at the interface location.  Rightmost frame (c) shows  plane-averaged profiles; the streamwise  component provides the interface slip length ($\langle L_{112}\rangle=0.15$ above $y_s$), whereas the average of wall-normal component is zero  (i.e. $\langle L_{212}\rangle=0$), since the 2D field is antisymmetric with respect to center axis.}
  \label{fig:ms-res-prof2}
\end{figure}

Returning to our 2D configuration with a flat interface and aligned coordinate system, $L_{ij1}$ are unforced problems, leading to trivial solution for all components. Out of the forced problems $L_{ij2}$,  the  components $L_{122}$ and $L_{222}$ are zero since the forcing is in a constrained direction, i.e., due to mass conservation, the motion in vertical direction is zero, when the no-slip condition is enforced at the bottom  boundary. We are thus left with only one non-trivial problem ($L_{i12}$), for which the flow fields are shown in Fig.~\ref{fig:ms-res-prof2}, along with corresponding plane averaged profiles.

\section{Effective interface conditions} \label{sec:eff-intf-cond}

This section provides the final forms of the effective boundary conditions by averaging the microscale solutions provided in the previous section. Since the interface cell is located at the boundary between free fluid and porous region, the conditions for
the free fluid should be evaluated using the solution above the interface, while conditions for the porous region should be evaluated using the solution below the interface.
In particular, to solve for the pore pressure Laplace equation (\ref{eq:sec2:model-gov-3}) below the interface, one needs a boundary condition for the pressure at the interface; this can be obtained by averaging the $\mathcal{O}(1)$-problem (see section~\ref{sec:mse-lead-ord-eq}). To solve the velocity of the free flow above the interface, one needs a condition for the velocity at the interface. We  recall however that the $\mathcal{O}(1)$-problem given by equations (\ref{eq:gov-mse-ord0-start}--\ref{eq:gov-mse-ord0-end}) does not contain velocity. Therefore, one has to investigate solution of the $\mathcal{O}(\epsilon)$-problem (equation \ref{eq:microsc-ansatz}) to determine a boundary condition for the velocity. This is a consequence of the pore velocity viscous term being higher order compared to the pressure gradient term, see $\ord$ pre-factors in equation (\ref{eq:gov-eq-fluid1}). 

\subsection{Condition for pore pressure}\label{sec:eff_press}

Using decomposition (\ref{eq:goveq-chan-decomp}) and the scaling for pressure perturbation (\ref{eq:press-pert-scale-above}), we can write the pressure field above the interface as
\begin{align}
p & = P + \ordest{\ord}.
\end{align}
Here, we have no angle brackets around $P$, since it is independent of the microscale coordinate.
Taking the average of the expression above in a microscale volume of size $l^3$ -- see definition (\ref{app-eq:def-average-vol}) -- above the interface gives,
\begin{equation}
\langle p \rangle =  P + \ordest{\ord}.
\label{eq:vol-avg-press-above}
\end{equation}
The multi-scale expanded  pressure field  (\ref{eq:ex22}) below the interface, on the other hand, is 
\begin{align}
p^{-} & = p^{-(0)} + \ordest{\ord},
\label{eq:approx-pressure-below}
\end{align}
where its volume averaged form is
\begin{align}
\langle p^{-} \rangle & =  p^{-(0)}  + \ordest{\ord}. \label{eq:vol-avg-press-below}
\end{align}
According to solution of the $\mathcal{O}(1)$-problem (\ref{eq:pressure-zero-sol-intf}),
one can state that the pressure field in the whole interface cell is constant and equal to the macroscale pressure $P$.
Inserting the solution of the $\mathcal{O}(1)$-problem (\ref{eq:pressure-zero-sol-intf})
in (\ref{eq:vol-avg-press-below}) and then equating it to expression (\ref{eq:vol-avg-press-above}),
we obtain the following at the interface,
\begin{align*}
\langle p^{-} \rangle & = \langle p \rangle + \ordest{\ord}. 
\end{align*}
For brevity, we denote averaged dimensional quantities with a ``hat''
(e.g. $\hat{p}^{-} = \Delta P \langle p^{-} \rangle$), 
which gives the pressure interface condition in its in final dimensional form,
%
\begin{align}
\hat{p}^{-} & = \hat{p}. \label{eq:approx-pressure-interf}
\end{align}
%
Working with the chosen estimates  (see Tab.~\ref{tab:scales}), one obtains pressure continuity up to $\ordest{\ord}$ for any anisotropic porous bed. We point out that from (\ref{eq:microsc-ansatz1}), one may formulate a pressure condition valid to $\mathcal{O}(\epsilon^2)$.
This is however out of the scope for this work.
We note that
this result is different compared to works by
\cite{marciniak2012effective} and \cite{carraro2013pressure}. This is a direct
consequence of the theoretical assumption $Re_f \sim \ord^{-1}$,
which leads to the $\ordest{1}$-problem for pressure being trivial.

\subsection{Velocity boundary condition for free fluid}\label{sec:5-2}

The solution to the  $\mathcal{O}(\epsilon)$-problem (see section~\ref{sec:mse-first-ord-eq}) is obtained numerically by computing $K_{ij}$ and $L_{ijk}$. One may then proceed to  construct the fully resolved flow field with error $\ordest{\ord}$ near the interface. First, we write the velocity above the interface as
\begin{align}
u_i^{+} & = u_i^{+(0)} + \ordest{\ord}.
\end{align}
The macroscale term $U_i$ does not appear in the expression above, because it is constant in the interface cell and this constant has be zero due to the boundary condition (\ref{eq:gov-eq-fluid1-chfast-bc}).
%
Inserting the above expression into (\ref{eq:microsc-ansatz}) gives
\begin{equation}
u_i^{+} =  - K^{+}_{ij} p^{-(0)}_{,j_0} + L^{+}_{ijk} \left( \left. U_{j,k} \right|_{y_s} + \left. U_{k,j} \right|_{y_s} \right) + \ordest{\ord}.
\end{equation}
We average out the microscale oscillations by forming the volume average above the interface,
\begin{equation}
\langle u_i^{+} \rangle =  - \langle K^{+}_{ij} \rangle\, p^{-(0)}_{,j_0} + \langle L^{+}_{ijk} \rangle\, \left( \left. U_{j,k} \right|_{y_s} + \left. U_{k,j} \right|_{y_s} \right)  + \ordest{\ord} .
\label{eq:def-interf-avg-field}
\end{equation}
Here, we have no angle brackets around pressure and velocity gradients, since these quantities are independent of the microscale coordinate.

Now, by inserting the approximations of the velocity strain (\ref{eq:vel-bc-str-relation}) and the pressure gradient (\ref{eq:vel-bc-pr-relation}) -- see Appendix~\ref{sec:vol-avg-relate} --  into  (\ref{eq:def-interf-avg-field}), we obtain
%
\begin{equation*}
\langle u_i \rangle  = - \mathcal{K}_{ij} \frac{1}{\ord} \langle p^{-} \rangle_{,j}
+ \mathcal{L}_{ijk} \left( \langle u_{j} \rangle_{,k} + \langle u_{k} \rangle_{,j} \right)  + \ordest{\ord},
\end{equation*}
where for convenience, we have denoted $\mathcal{K}_{ij} = \langle K^{+}_{ij} \rangle$ and $\mathcal{L}_{ijk} = \langle L^{+}_{ijk} \rangle$. The coefficients $\mathcal{K}_{ij}$ and $\mathcal{L}_{ijk}$ are evaluated over $l^3$ cube ($l^2$ in 2D setting) at the top of the finite interface cell.
Finally, we have used that $u_i = U_i + u^{+}_i = u^{+}_i$ at the interface.
%

In order to return to the boundary conditions (\ref{eq:sec2:effec-v}--\ref{eq:sec2:effec-u}) used for the lid-driven cavity problem in section \ref{sec:dns-model-cavity}, we revert the boundary condition to dimensional quantities,
\begin{equation}
\hat{u}_i = - \left( \mathcal{K}_{ij} \frac{l^2}{\mu} \right) \hat{p}^{-}_{,j} +  \left( \mathcal{L}_{ijk} l  \right) \left( \hat{u}_{j,k} + \hat{u}_{k,j} \right). \label{eq:model-eq-intf-bc-general-dim}
\end{equation}
%
The coefficients, based on solutions of the interface cell problems, for the cavity flow are 
\[
K_{cyl} = l^2\,  \mathcal{K}_{22},\quad K_s = l^2\, \mathcal{K}_{11}\qquad \textrm{and}\qquad L_s =  l\, \mathcal{L}_{112}. 
\]
All other components of tensors $\mathcal{K}_{ij}$ and $\mathcal{L}_{ijk}$ are zero for the porous bed with regular circular cylinders, and therefore there is no velocity shear term for the penetration velocity (\ref{eq:sec2:effec-v}). Actual values used in section~\ref{sec:dns-model-cavity} are $\mathcal{K}_{11} = 0.0312$, $\mathcal{K}_{22} = 0.0986$, and $\mathcal{L}_{112} = 0.1783$.

Equation (\ref{eq:model-eq-intf-bc-general-dim}) is the final expression of the velocity  boundary condition for a rigid porous bed. It can be used together with Navier-Stokes equations in any domain of interest, in order to take into account the effects of the porous medium, without resolving the microscale flow within the porous bed. We emphasize that the ``minus'' notation for pressure means that the pressure gradient in the boundary condition is the gradient of the pore pressure. In the final subsection, we test the robustness of the derived boundary condition by varying solid volume fraction, scale separation parameter and interface location.

\begin{table}
  \begin{center}
 \begin{tabular}{ccc | llll }
 $\ys$ & $\ord$ & $\phi$ & \multicolumn{1}{c}{ $\hat{u}_{sK} / \bar{u}_s$ } & \multicolumn{1}{c}{$\hat{u}_{sL} / \bar{u}_s$} & \multicolumn{1}{c}{$\hat{u}_s / \bar{u}_s$}  & \multicolumn{1}{c}{$\bar{u}_s / \bar{u}_s$}  \\ 
 \multirow{6}{*}{ $ 0.10 $ } &  \multirow{3}{*}{ $ 0.02 $ } & $0.02$ & $1.61 \cdot 10^{-2}$  & $9.70 \cdot 10^{-1}$ & $9.87 \cdot 10^{-1}$ & $1.0$ \\
  & & $0.15$ & $8.32 \cdot 10^{-3}$ & $9.85 \cdot 10^{-1}$  & $9.94 \cdot 10^{-1}$ & $1.0$ \\
  & & $0.45$ & $7.67 \cdot 10^{-3}$ & $9.87 \cdot 10^{-1}$  & $9.95 \cdot 10^{-1}$ & $1.0$  \\
  &  \multirow{3}{*}{ $ 0.10 $ } & $0.02$ & $7.02 \cdot 10^{-2}$ & $9.36 \cdot 10^{-1}$ & $1.02 \cdot 10^{0}$ & $1.0$ \\
  & & $0.15$ & $3.81 \cdot 10^{-2}$ & $9.57 \cdot 10^{-1}$ & $1.00 \cdot 10^{0}$ & $1.0$ \\
  & & $0.45$ & $3.56 \cdot 10^{-2}$ & $9.59 \cdot 10^{-1}$ & $1.00 \cdot 10^{0}$ & $1.0$  \\
  \hline
   \multirow{6}{*}{ $ 0.50 $ } &  \multirow{3}{*}{ $ 0.02 $ } & $0.02$ & $2.89 \cdot 10^{-2}$  & $9.69 \cdot 10^{-1}$ & $9.99 \cdot 10^{-1}$ & $1.0$  \\
  & & $0.15$ & $2.60 \cdot 10^{-2}$ & $9.73 \cdot 10^{-1}$ & $1.00 \cdot 10^{0}$ & $1.0$  \\
  & & $0.45$ & $2.55 \cdot 10^{-2}$ & $9.74 \cdot 10^{-1}$ & $1.00 \cdot 10^{0}$ & $1.0$  \\
  &  \multirow{3}{*}{ $ 0.10 $ } & $0.02$ & $1.38 \cdot 10^{-1}$ & $9.19 \cdot 10^{-1}$ & $1.06\cdot 10^{0}$ & $1.0$ \\
  & & $0.15$ & $1.29 \cdot 10^{-1}$ & $9.19 \cdot 10^{-1}$ & $1.05 \cdot 10^{0}$ & $1.0$  \\
  & & $0.45$ & $1.28 \cdot 10^{-1}$ & $9.16 \cdot 10^{-1}$ & $1.04 \cdot 10^{0}$ & $1.0$
  \end{tabular}
    \caption{Cavity slip velocity values at maximum predicted by model $\hat{u}_s$ normalized with DNS result $\bar{u}_s$. The DNS results are plane averaged over length of microscale. Darcy contribution $\hat{u}_{sK} = -K_s / \mu\, \pd_x \hat{p}^{-}$ and Navier slip contribution $\hat{u}_{sL} = L_s ( \pd_y \hat{u} + \pd_x \hat{v} )$ are listed separately.}
  \label{tab:chan-slip-comp}
  \end{center}
\end{table}

\subsection{Accuracy of slip prediction and robustness to interface location}\label{sec:accuracy}
We now return to the example of the lid-driven cavity with a porous bed in order to illustrate more quantitatively that the proposed boundary condition yields accurate and robust slip velocity predictions. 
More specifically, we carry out a parametric study and report predictions of maximum slip velocity $\hat{u}_s$ at the center of the cavity. In order to do a fair comparison, we surface average the DNS results $\bar{u}_s = \langle u_s \rangle_S$ at the interface.
We also assess the contributions from two different terms in the derived boundary condition (\ref{eq:model-eq-intf-bc-general-dim}). Tab.~\ref{tab:chan-slip-comp} shows that for the range of parameters considered,  the contribution at the interface from the Navier-slip term is always at least an order of magnitude larger than the contribution from the Darcy term. This is a consequence of (\ref{eq:model-eq-intf-bc-general-dim}), where $K_s \sim {l^2}$ and $L_s \sim {l}$, and therefore $K_s \ll L_s$ for fine microstructures.
This result is in agreement with previous work. It was first suggested by \cite{saffman1971boundary} that the Darcy term is of higher order and can be neglected, the same result was later rigorously proved by \citet{mikelic2000interface}.
One may therefore obtain a good approximation with only the Navier-slip term, as first suggested by  \citet{saffman1971boundary} and later rigorously shown by \citet{mikelic2000interface}.  Including the Darcy  term however yields -- consistently -- a smaller error, and therefore also   a \textit {robust} velocity boundary condition with respect to the interface location and different pore geometries. Additionally, the Darcy term is the only contribution appearing in the interface normal direction, which is essential to capture the momentum transfer from and to the porous region.

Note that although the Darcy term is much smaller than the Navier slip term, both terms appear in the $\mathcal{O}(\epsilon)$ equation (see section~\ref{sec:mse-first-ord-eq}).
This is a consequence of the estimate that perturbation velocity is of the same order as the Darcy velocity, i.e.~$u^{\pm}_i \sim U^d$ and that the fast flow velocity scales as equation (\ref{eq:assumption-on-Uf}). An alternative approach would be to assume $U^{f} \sim \ord^{-2} U^d$, which would essentially result in three velocity scales to allow for the slip velocity to be faster than the Darcy flow but slower than the free flow. In such an approach -- which would more sophisticated than the current one -- the Darcy term can appear in higher-order equation than the equation for which the Navier-slip term appears. Nevertheless, one can observe that our direct numerical simulations are in good agreement with simulations of the homogenized model despite that we strictly speaking do not satisfy (\ref{eq:assumption-on-Uf}) for Stokes flow. Thus our method can be considered as practical parameter-free framework for computing the coefficients of a generalized condition proposed by Beavers and Joseph; a condition which has been employed under a variety of different flow conditions by experimentalists.
To sum up, we have theoretically
assumed that (i) $\ord \ll 1$; (ii) $Re_d \leq 1$ and (iii) $Re_f \sim \ord^{-1}$.
Based on numerical tests in this section
for $Re_f = 0$ and up to $\ord = 0.1$, we have determined the
practical limits of the derived interface condition by relaxing
conditions (i) and (iii), as summarized
in section~\ref{sec:mod-desc}.

\section{Conclusions} \label{sec:conclusions}
We have presented a framework to construct a reduced homogenized model of the flow above and through a porous medium consisting of regular solid structures of general shape.
The main contribution of the present paper is to provide the foundation and the tools to compute effective boundary conditions completely free from data fitting. The approach that we adopted can be summarized by the following four steps. First, the governing equations are made non-dimensional using scale estimates arising from flow in the porous domain. Second, the governing equations describing the fully resolved flow are separated at a virtual interface, and decomposed above the interface into equations for the fast flow and for perturbations. Third, multi-scale expansion according to \cite{mei2010homogenization} is employed on perturbation and pore equations. Finally and after solving $\ordest{1}$- and $\ordest{\ord}$-problems, we construct the interface conditions for pore pressure and fluid velocity using volume averages.

This procedure results in macroscopic description of the flow over a porous domain with an error $\ordest{\ord}$. Specifically, using our a priori scaling estimates, the leading order conditions are a  pressure continuity condition and a generalized tensorial BJ boundary condition. The  proposed velocity condition depends on the \textit{interface} permeability and on the velocity strain, while the BJ condition contains \textit{interior} permeability and velocity derivative of one component in one direction only. To the authors' knowledge, such a general formulation has not been derived and validated before. Moreover, in order to obtain the constants of the effective boundary conditions, we derive a number of Stokes problems that need be solved numerically in small interface unit cells.  Solvers for microscale problems have been released as an open-source software \citep{github2016UgisShervin}, along with solver used for the lid-driven cavity flow.

This work is also among the first to validate non-empirical boundary conditions on 2D flows with DNS, where penetration velocity, slip velocity and pressure condition have to be predicted to solve the coupled two-domain problem. The present boundary condition has been tested in the lid-driven cavity flow for a range of volume fractions from $\phi = 0.02$ to $0.45$, scale separation parameters from $\ord = 0.02$ to $0.1$ and interface locations from $y_s = 0.1$ to $y_s = 0.5$. When the homogenized model results are  compared to DNS, the  slip velocity predictions have been found to be robust and to give accurate predictions for all investigated parameters. 
We hope that, with this work and the release of the associated software, we can provide the numerical fluid dynamics community the tools to model flows over existing non-smooth surfaces as well as to design surfaces to modify fluid flow characteristics. 

\vspace{-0.3cm}
\section*{Acknowledgements}
U.L and S.B. acknowledges the financial support from the Swedish Research Council (VR-2014-5680). We also thank Prof. Alessandro Bottaro,  Dr. Giuseppe Antonio Zampogna and Dr. Sudhakar Yogaraj for fruitful discussions.

\appendix

\section{Momentum balance and order estimates}\label{sec:app-scale-estim}
This appendix provides a description of the physical scales in the porous medium and in the free fluid region, which are used in section \ref{sec:gov-decomp-problem} and summarized in Tab.~\ref{tab:scales}. For a dense coating, where inertial effects can be neglected, the momentum balance in the porous region at the pore scale is
\[
\mu \nabla_p^2 {\bf u} - \nabla P -\nabla_p p= 0,
\]
where the viscous force is balanced by the sum of the macroscopic pressure driving the flow and the microscopic pressure at the pore scale. Here, $\nabla_p=()_{,j_1}$ denotes the gradient at the pore scale and $\nabla=()_{,j_0}$ denotes the gradient at the macroscale.  This is a classical result at first order in the interior, as derived by \cite{mei2010homogenization} and also used by \cite{gopinath2011elastohydrodynamics}. The force balance is thus
\begin{equation}
\frac{\mu U^{d}}{l^{2}}  \sim { \frac{\Delta P}{H} } \sim \frac{\Delta p}{l},
\label{eq:scaling2}
\end{equation}
where  $U^d$ is the characteristic velocity in the porous region  and $\Delta p$ is characteristic microscopic pressure. 

Comparing first and second term, we immediately arrive with the first estimate used in the main paper (equation \ref{eq:first-estim}). We argue in the main paper, that the perturbation velocities are caused by porous medium and therefore could be estimated based on characteristic velocity in the porous region
\begin{equation}
\tilde{u}^{\pm}_i \sim U_d. \label{eq:app:variable-estimate-start}
\end{equation}
The pore pressure is associated with the global pressure difference and estimated as
\begin{equation}
\tilde{p}^{-} \sim \Delta P,
\end{equation}
whereas the pressure perturbation above the interface we associate with the microscopic pressure difference $\Delta p$. We argue that the pressure perturbation above is caused directly by the flow in the pores, which results in pressure difference in the pore scale. The estimate we use is
\begin{equation}
\tilde{p}^{+} \sim \Delta p.
\end{equation}
%
%
For the fast flow, we use an estimate
\begin{equation}
\tilde{U}_i \sim U^f,
\end{equation}
where $U^f$ is the characteristic fast flow velocity. The associated pressure we estimate using the same macroscopic pressure difference as everywhere else
\begin{equation}
\tilde{P} \sim \Delta P. \label{eq:app:variable-estimate-end}
\end{equation}


The magnitude of the fast flow velocity $U^f$ (\ref{eq:assumption-on-Uf}) 
can be estimated from momentum balance in the free fluid.
The momentum of the fast flow is governed by
\[
 \rho_f \left( \pdt {\bf U} + \left( {\bf U} \cdot \bnabla \right) {\bf U} \right)
  = \mu \nabla^2 {\bf U} - \nabla P.
\]
This provides a balance between pressure gradient, viscous force and inertial effects,
\begin{equation}
\frac{\Delta P}{H} \sim \frac{\rho_f \left( U^{f} \right)^2}{H}\sim\frac{\mu U^f}{H^2}.
\label{eq:scaling1}
\end{equation}
As mentioned in the main text, there is no unique way perform a priori estimates. One  approach is to use similarity between global pressure gradient and inertial term. Combining that with balance between global pressure gradient and seepage flow (\ref{eq:scaling2}) gives
\begin{equation}
\frac{\rho_f \left( U^{f} \right)^2}{H} \sim \frac{\Delta P}{H} \sim \frac{\mu U^{d}}{l^{2}},
\end{equation}
which after rearranging can be written as
\begin{equation}
U^f \sim \frac{1}{Re_f} \frac{1}{\ord^2} U^d,
\end{equation}
where $Re_f = \rho_f H U^f / \mu$ is free fluid Reynolds number. Now one has to say something about $Re_f$ in order to finalize estimate of $U^f$. One possible choice is to assume $Re_f = \ordest{\ord^{-1}}$, and then we recover assumption (\ref{eq:assumption-on-Uf}). The estimate of $U^f$ is later implicitly
used in order to determine the order
of shear stress from the free fluid at the interface
\begin{equation}
\tilde{U}_{i,j} \sim \frac{U^f}{H}, \label{eq:app:variable-estimate-shear}
\end{equation}
where we have assumed that the fast free flow velocity is obtained over the macroscopic
length scale.

Finally, estimates (\ref{eq:app:variable-estimate-start}--\ref{eq:app:variable-estimate-end}) can be made non-dimensional following section (\ref{eq:nondim}). After using momentum balance presented here (\ref{eq:scaling2}) and assumption  (\ref{eq:assumption-on-Uf}), one arrives with dimensionless orders presented in right-most column of Tab.~\ref{tab:scales}. Additionally, making the shear stress estimate
(\ref{eq:app:variable-estimate-shear})
non-dimensional, one obtains $U_{i,j} = \ordest{1}$, which together with $\ord$
pre-factor in the non-dimensional stress (\ref{app-eq:def-stress3}) leads to
free fluid shear
appearance in $\ordest{\ord}$-problem (\ref{eq:gov-microcell-fluid4}).

\section{Definitions}\label{sec:app-def}
Here we provide definitions of various tensors and operators used in the main paper. We start with linear Stokes operator
\begin{align}
\Ac \left(u_{i}, p, \ord \right)  = R_i : & & -p_{,i}+\ord u_{i,jj} & = R_i, \label{app-eq:def-Stokes} \\
& & u_{i,i} & = 0, \nonumber \\
& & \left . u_{i} \right|_{\Gamma_c} & = 0, \nonumber
\end{align}
where $R_i$ is a right-hand term, usually containing the inertial terms. A similar operator is used for microscale problems within the multi-scale expansion
\begin{align}
\Ac_1 \left(u_{i}, p, 1 \right)  = R_i : & & -p_{,i_1}+ u_{i,j_1j_1} & = R_i, \label{app-eq:def-Stokes-micro} \\
& & u_{i,i_1} & = 0, \nonumber \\
& & \left . u_{i} \right|_{\Gamma_c} & = 0, \nonumber
\end{align}
where derivative with respect to microscale $()_{,i_1}$ is defined in equation (\ref{eq:mse-chain-der}).
Fluid stress tensors used in this paper are
\begin{align}
\Sigma^{u^+}_{ij} & = -p^{+} \delta_{ij} + \ord \left( u^{+}_{i,j} + u^{+}_{j,i} \right), \label{app-eq:def-stress1} \\
\Sigma^{u^-}_{ij} & = -p^{-} \delta_{ij} + \ord \left( u^{-}_{i,j} + u^{-}_{j,i} \right), \label{app-eq:def-stress2} \\
\Sigma^{U}_{ij} & = -P \delta_{ij} + \ord \left( U_{i,j} + U_{j,i} \right), \label{app-eq:def-stress3}
\end{align}
for the flow field, and
\begin{align}
\vare[u^+]{1}{\Sigma_{ij}} & =-\vare[+]{1}{p}\delta_{ij}+(\vare[+]{0}{u_{i,j_1}}+\vare[+]{0}{u_{j,i_1}}), \\
\vare[u^-]{1}{\Sigma_{ij}} & =-\vare[-]{1}{p}\delta_{ij}+(\vare[-]{0}{u_{i,j_1}}+\vare[-]{0}{u_{j,i_1}}),
\end{align}
for the $\ordest{\ord}$-problem in microscale. Stress tensors for permeability interface problem are
\begin{align}
\Sigma^{K^+}_{ij} & =-A^{+}_k \delta_{ij}+(K^{+}_{ik,j_1}+K^{+}_{jk,i_1}), \\
\Sigma^{K^-}_{ij} & =-A^{-}_k \delta_{ij}+(K^{-}_{ik,j_1}+K^{-}_{jk,i_1}),
\end{align}
and stress tensors for Navier-slip interface problem are
\begin{align}
\Sigma^{L^+}_{ij} & =-B^{+}_{kl} \delta_{ij}+(L^{+}_{ikl,j_1}+L^{+}_{jkl,i_1}), \\
\Sigma^{L^-}_{ij} & =-B^{-}_{kl} \delta_{ij}+(L^{-}_{ikl,j_1}+L^{-}_{jkl,i_1}).
\end{align}
The velocity strain tensor is
\begin{equation}
S_{ij} = U_{i,j}+U_{j,i}.
\end{equation}
In order to homogenize the results, we use a volume average operator
\begin{equation}
\langle f \rangle \left(  \vec{X} \right) = \frac{1}{l^3} \int\limits_{-l/2}^{l/2} \int\limits_{-l/2}^{l/2} \int\limits_{-l/2}^{l/2} f \left( \vec{x} - \vec{X}\right) \,\mathit{dx}\,\mathit{dy}\,\mathit{dz}, \label{app-eq:def-average-vol}
\end{equation}
where $f$ is some variable and $\vec{X}$ is the location
of the center of the averaging volume.
In two 
dimensional case, this integral reduces to a surface integral. For investigations of 
interface cell results, we use a surface average operator
\begin{equation}
\langle f \rangle_S \left(  \vec{X} \right) = \frac{1}{l^2}\int\limits_{-l/2}^{l/2} \int\limits_{-l/2}^{l/2} f \left( \vec{x} - \vec{X}\right) \,\mathit{dx}\,\mathit{dy}, \label{app-eq:def-average-surf}
\end{equation}
where plane is oriented parallel to the interface. Here, $\vec{X}$ is the location of the center of the averaging surface. In two dimensional case as in the lid-driven cavity example reported in the  paper, this integral reduces to a line integral. For the convenience, the $\vec{X}$ argument is omitted in the main paper.

\section{Homogenized velocity strain and pressure gradient} \label{sec:vol-avg-relate}

In this appendix we derive two expressions for the volume averaged pressure gradient in the porous region and the volume-averaged velocity gradient in the free flow region. These expressions  are used in section \ref{sec:5-2}. 

We start by forming the average of the velocity (\ref{eq:goveq-chan-decomp}) in a microscale volume of size $l^3$, as defined in equation (\ref{app-eq:def-average-vol}),
\begin{equation}
\langle  u_{j} \rangle = U_{j} + \langle  u_{j}^{+(0)} \rangle + \ordest{\ord}. \label{eq:app-strainterm-avg-start}
\end{equation}
The averaging volume can be centered at any $y-$coordinate as long as it is within the free fluid part of the interface unit cell. There are no brackets around velocity $U_{j}$ because it does not depend on the microscale coordinate, i.e., it is constant over length $l$. Here we have also used that $u_j^{+} = u_j^{+(0)} + \ordest{\ord}$. Next, we take the derivative of the expression above and use the chain rule (\ref{eq:mse-chain-der}), 
\begin{equation}
\langle  u_{j} \rangle_{,k} = U_{j,k} + \langle  u_{j}^{+(0)} \rangle_{,k_1} + \ordest{\ord},
\end{equation}
where all the derivatives with respect to $X_i$ appear in $\ordest{\ord}$ term due to the $\ord$ pre-factor in the chain rule (\ref{eq:mse-chain-der}). Additionally, the microscale dependence of the term $u_{j}^{+(0)}$ is averaged out, therefore we conclude that $\langle  u_{j}^{+(0)} \rangle_{,k_1} = 0$.
Finally we arrive with relationship between fast flow velocity gradient and the averaged free flow velocity gradient, 
\begin{equation}
\langle  u_{j} \rangle_{,k} = U_{j,k} + \ordest{\ord}. \label{eq:vel-bc-str-relation}
\end{equation}
This expression is defined anywhere in the free fluid domain, including at the interface with the porous region.

Next, we  average the pressure expansion (\ref{eq:ex22}) in a $l^3$ volume centered at an arbitrary point in the porous domain, which after taking the gradient, results in
\begin{align}
\langle p^- \rangle_{,j}  = p^{-(0)}_{,j} + \ord \langle p^{-(1)} \rangle_{,j} + \ordest{\ord^2} = \ord p^{-(0)}_{,j_0} + \ordest{\ord^2}, \label{eq:vel-bc-pr-relation}
\end{align}
where we have used that $\langle p^{-(1)} \rangle_{,j_1} = 0$. 
The macroscale derivative of $\langle p^{-(1)} \rangle$ is again absorbed in
$\ordest{\ord^2}$ terms due to the additional $\ord$ pre-factor. For the
leading order pressure we have used the chain rule (\ref{eq:mse-chain-der}) and
the fact that it is independent of microscale
(\ref{eq:gov-mse-ord0-end}).
This expression is
defined anywhere in the porous domain, including at the interface with the free fluid.

\bibliographystyle{jfm}
\bibliography{/home/ugis/Documents/references_UgisL,refs}
\end{document}